# Crystallisation and Local Order in Glass-Forming Binary Mixtures

Julián R. Fernández[*] and Peter Harrowell

*School of Chemistry, University of Sydney, New South Wales, 2006, Australia*
*and*
[*]*Comisión Nacional de Energía Atómica, Av. Libertador 8250*
*Capital Federal, Buenos Aires, Argentina*

**Abstract.** The local organisation of a simulated glass-forming mixture due to Kob and Andersen is analysed. Evidence is presented for a structural transition from triangulated coordination polyhedra to cubic as the number fraction of the smaller species B increases towards equimolar. The impact of the change on the partial radial distribution function $g_{BB}(r)$ is established. The dependence of the crystallisation rate on the composition is determined and the related to the observed structural changes.

## INTRODUCTION

Mechanical stability confers a special status on the associated configuration. Such special structures, no matter how apparently 'disordered', are subject to configurational restrictions by virtue of their stability. The 'deeper' the stability, the greater the configurational constraint, with long-range order (crystallline or orientational only) representing possible endpoints of this reduction in configuration space. The question of amorphous structure corresponds to establishing just how these constraints are manifested for the given stable configurations and whether these constraints provide a useful means of quantifying a type of order.

In this paper we present a preliminary study of the structure of the amorphous states of a popular model of a glass-forming alloy based on a binary mixture of particles interacting via Lennard-Jones (LJ) potentials. This work extends our recent study [1,2] of the stable and metastable crystalline phases of LJ mixtures.

The literature on amorphous structures predates the development of modern liquid theory and, therefore, may not be familiar to all readers. For this reason we have included a brief and, unavoidably, subjective summary of the history of this topic. We then introduce the model mixture and algorithms. Our results are divided into four sections. First, we examine the nature of coordination geometry about the minority component. We then present a simple argument which establishes a relationship between the composition and the average number of coordination polyhedra that must be packed about each majority component. This constraint is then used to rationalise the anomalous behaviour observed in the radial distribution function and the crystallisation rates.

## A VERY BRIEF HISTORY OF AMORPHOUS STRUCTURE

A persistent dichotomy between *homogeneous* and *heterogeneous* pictures of glass structure can be traced back at least as far as the 1930's. In 1932 Zacharaisen [3] described a random network some 30 years before Bernal [4] and Finney [5] gave it substance in their packing studies of hard spheres. It was Bernal, in particular, who vigorously asserted the homogeneous character of such random packing. The random close packed model was extended to amorphous metal-metalloid alloys by Polk [6] but failed to account for the observed variation in metal-metal scattering with the choice of the metalloid [7]. The idea of the homogeneous random network has proved most useful in the low coordinated glasses such as silica and the chalcogenide mixtures. The constraint theory of Phillips [8] and Thorpe [9] provides a powerful set of limits on the stability of homogeneous random networks.

The heterogeneous picture of glass structure found an early proponent in Tammann [10]. In contrast with Zacharaisen's *structural* prescription, Tammann pictured the glass forming *process* as analogous to that of clays in which rigid clusters gradually come into contact as the intervening liquid is removed. Most of the efforts to 'flesh out' this idea begin with the suggestion of Frank [11] in 1952 that the stability of liquids to crystallisation might be due to the stability of icosahedral clusters. This idea has since been considerably extended in both application and sophistication [12]. Hoare [13] has presented one of the most lucid accounts of the program to demonstrate that Tammann's clusters could be constructed out of clusters characterised by 5- and 10-fold symmetry. Hoare, in particular, has emphasised the importance of explaining how the growth of these 'rigid clusters' comes to be self-limiting. An alternative, non-structural, perspective on glass formation has developed over the last 10 years that presents a picture very much in the spirit of Tammann's idea. This approach is based on the measurement of the growth of dynamic heterogeneities with supercoolong.

Given the perennial popularity of the idea of icosahedral organisation in the context of amorphous structure it is worth making three points. First, the low energy of isolated icosahedral clusters of a single spherical species arises from their low surface energy and is not reproduced in condensed phases except in the case of oscillatory interactions such as those examined by Dzugutov [14]. Second, while the icosahedron cannot uniformly fill space it certainly forms stable crystals and, therefore, its presence alone is insufficient to explain the absence of crystallisation. Finally, of the 110 convex polyhedra with regular faces (excluding the prisms and antiprisms) only 3 can uniformly fill space. The icosahedron, therefore, must merely take its place among the many local coordination geometries that might frustrate crystallisation. Gaskell [15], for example, has provided strong evidence for the important role of a 9-coordinated polyhedra, the tricapped trigonal prism, in the amorphous Ni-P alloys. The successful packing criteria developed by Egami [16] to identify glass-forming alloys also allows for a wide range of coordination polyhedra.

This, then, is the point from which we shall start. Our initial perspective on amorphous structure is local (i.e nearest neighbour), as determined by the short range of the particle interactions. We shall, however, be led to consider the intermediate range, i.e. those lengths that cover the packing of adjacent coordination polyhedra. We shall argue that is unlikely that structure over any longer length scales is necessary to stabilise an amorphous state over the limited run times and system sizes accessible computationally.

## MODEL AND ALGORITHM

The Lennard-Jones (LJ) potential for a mixture has the form

$$V_{ij}(r) = 4\varepsilon_{ij}\left[\left(\frac{\sigma_{ij}}{r}\right)^{12} - \left(\frac{\sigma_{ij}}{r}\right)^{6}\right] \quad (1)$$

where the sub-indices *i* and *j* could take the values A or B. We truncate the potential at a distance $2.5\sigma_{ij}$ and shift the potential so that it equals zero at the cut-off. (Here we shall set the masses of both components equal to *m*.) We shall work in the following reduced units throughout this paper: the unit of length is $\sigma_{AA}$, the unit of energy $\varepsilon_{AA}$, and the unit of time $\tau = \sigma_{AA}(m/\varepsilon_{AA})^{1/2}$. We shall follow Kob and Andersen (KA) [17] and set $\sigma_{AB} = 0.8$, $\sigma_{BB} = 0.88$, $\varepsilon_{AB} = 1.5$ and $\varepsilon_{BB} = 0.5$. This mixture at a composition of $x_B = N_B/N = 0.2$ has been studied extensively as a model glass former. The choice of parameters was originally made to model the Ni-P system. Based on the stable crystal phases [1], we have suggested that it better represents the Ni-Be mixture.

Molecular dynamics simulations have been carried out at constant NPT using a Nosé-Poincaré-Andersen Hamiltonian and a generalised leapfrog algorithm [18]. All calculations were performed at zero pressure. Enthalpy minimizations were carried out using a conjugate gradient scheme which ensures a fixed pressure.

# RESULTS

## The Stable and Metastable Crystal Structures

To establish a well-defined point of departure, we shall begin with the stable and metastable crystal structures of the KA mixture. We have recently completed a study of the lattice energies of a range of LJ mixtures [2]. In Figure 1 we plot the lattice energies per particle of a number of crystal structures as a function of $\sigma_{AB}$ for $A_3B$ mixtures (i.e. $x_B = 0.25$).

As reported previously, the most stable crystal state of the KA mixture over the composition range $0.0 \leq x_B \leq 0.5$ consists of coexisting face centered cubic (fcc) of pure A and the CsCl structure with composition AB. In order of ascending lattice energies we have the $PuBr_3$ structure, coexisting $Pd_2Zr$ structure and fcc, the cementite $Fe_3C$ structure and then the $Ni_3P$ structure. For reference, the lowest 'lattice energy' per particle obtained for the amorphous state following a similar enthalpy minimization is -7.92, significantly higher than the lattice energy of any of the crystalline states identified.

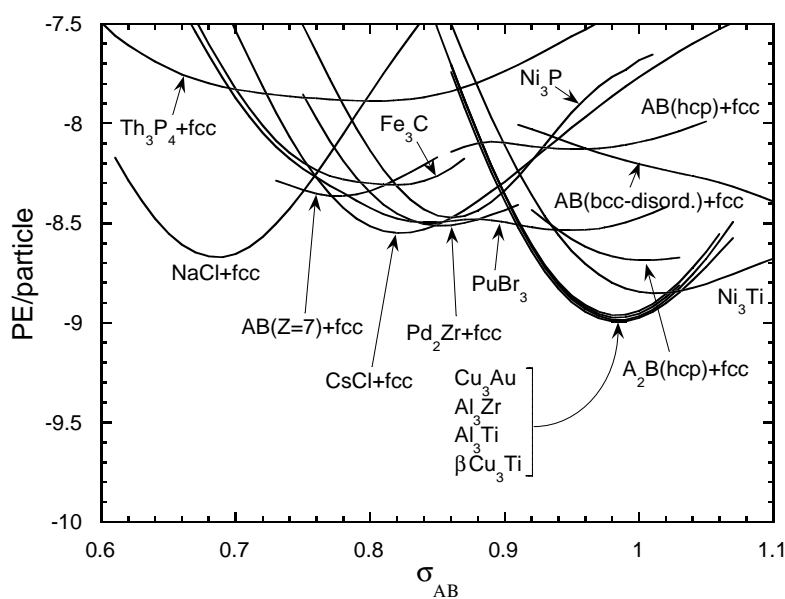

**FIGURE 1.** Lattice enthalpies per particle vs $\sigma_{AB}$ of a number of binary crystal structures at a composition $x_B = 0.25$. Note that the KA mixture corresponds to $\sigma_{AB} = 0.8$.

In the CsCl structure each B particle lies in the center of a cube of eight A particles. In the $PuBr_3$, $Fe_3C$ and $Ni_3P$ structures, each B particle lies in the center of a tricapped trigonal prism (see Figure 4) consisting of nine A particles.

## The Coordination of B Particles in the Amorphous State

In the KA mixture, the A-B interaction is strongly favoured over the B-B interaction. As a result, for $x_B < 0.5$ each B particle has only A nearest neighbours at low temperatures. In Figure 2 we plot the fraction of B particles with 7, 8 or 9 A neighbours for the amorphous mixture with $x_B = 0.25$ as a function of temperature. Note that there is a significant systematic change in the local coordination of the B particles on cooling with 8- and 9-fold clusters dominating at low temperatures. The low temperature limit of the B coordination has not yet been established.

As each B particle represents the centre of a polyhedron of A particles, then it follows that any two B particles that share four A neighbours represent two polyhedra sharing a 4-fold face. Similarly, two B particles that share three A neighbours correspond to adjacent polyhedra sharing a triangular face, and so on. We shall refer to such B particles pairs as "$B^nA$ bonds" where *n* is 4, 3, 2 or 1 depending on the number of A neighbours shared by the two B particles.

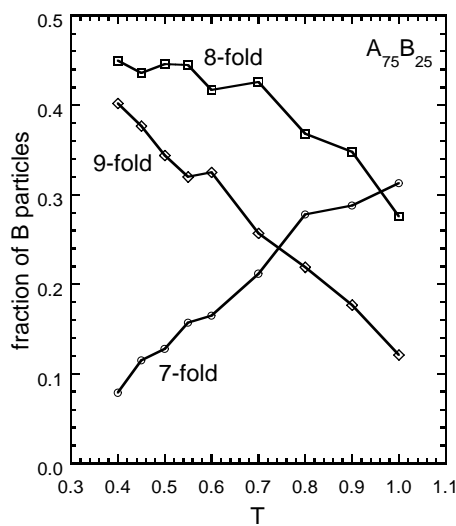

**FIGURE 2**. The fraction of B particles with 7, 8 or 9 A neighbours as a function of temperature in a $x_B = 0.25$ mixture.

This identification of $B^4A$ and $B^3A$ bonds with shared faces of the coordination polyhedra allows us to establish a picture of the geometry of the B coordination. By way of example, we plot in Figure 3 the distribution of angles between pairs of $B^3A$ bonds about individual B particles with 9-fold coordination in the $x_B = 0.25$ mixture at $T = 0.4$. We find that the peaks in this distribution correspond closely with the angles between surface normals of the perfect tricapped trigonal prism and conclude that that this geometry provides a reasonable description of the 9-fold coordination. The 8-fold coordination in the amorphous state is more ambiguous and it appears to be best described as a mixture of geometries dominated by the triangular dodecahedron and the square antiprism.

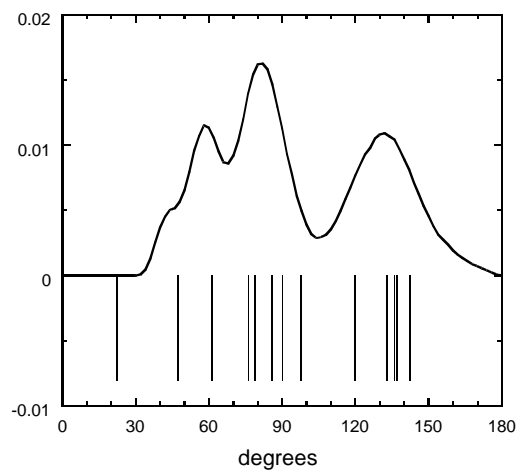

**FIGURE 3**. The distribution of angles (curve) between pairs of $B^3A$ bonds about individual B particles with 9-fold coordination in an $x_B = 0.25$ mixture at $T = 0.4$. The vertical lines represent the angles between surface normals of the tricapped trigonal prism.

Illustrations of the triangular dodecahedron (TD) and the tricapped trigonal prism (TTP) are provided in Figure 4. For the discussion that follows it is worth emphasising that both of these polyhedra have only triangular faces, in contrast to the square faces associated with B coordination in the stable CsCl crystal. It is also worth noting that while neither of these polyhedra can uniformly fill space both do occur in stable crystals; $Th_3P_4$ in the case of the TD polyhedron and, in the case of the TTP polyhedron, a number of crystals including $Fe_3C$, as already mentioned.

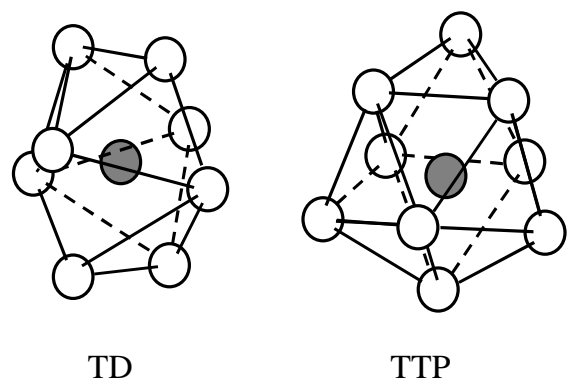

**FIGURE 4**. Illustrations of the triangular dodecahedron (TD) and tricapped trigonal prism (TTP).

## Packing Constraints for Polyhedra

Having characterised the coordination polyhedra, we need to consider how they pack in space. In spite of the considerable interest in the sphere packing problem, there is surprisingly little known about the packing of polyhedra. Here we present a simple argument that relates the composition and average number of B-centered polyhedra that share a vertex (i.e. an A particle). The argument assumes homogeneity of composition.

Let the average coordination of a B particle by A particles only by $\eta_B$. Then the average number of AB bonds is $N_B \eta_B = N_A \eta_A$, where $\eta_A$ is the average number of B neighbours about each A particle. We thus have

$$\eta_A = \eta_B N_B/N_A = \eta_B x_B/(1-x_B) \qquad (2)$$

In the previous section we established that at low temperatures $\eta_B$ is either 8 or 9 and, according to Eq. 1, $\eta_A$ must be less than that as long as $x_B < 0.5$. As each B particle lies at the centre of a polyhedron of A's, $\eta_A$ is equal to the number of polyhedra that share the vertex occupied by that A particle. There must be a geometrical limit as to how many polyhedra can physically share a vertex. For some guidance, lets look at this number in some crystals. In $Fe_3C$ with TTP coordination polyhedra, each A particle has only 3 B neighbours. In $Al_2Cu$ with the square antiprism coordination, four of these polyhedra can meet at any vertex. In $Th_3P_4$, six TD coordination polyhedra meet about any A particle. The maximum possible packing of polyhedra goes to the cubic coordination found in the CsCl structure in which 8 B particles surround each A particle. At this stage these values represent our best estimates of the limit of packing for each type of polyhedron in an amorphous phase. We have, of course, not considered the case of mixed polyhedra.

As $\eta_A$ increases with composition, this geometrical constraint (whatever it turns out to be) must be met, resulting, ultimately, in the exclusion of all coordination polyhedra except cubic. We find that the properties associated with the amorphous states with composition $x_B$ close to equimolar include rapid crystallisation and relatively low diffusion constants.

## The Anomalous Behaviour of $g_{BB}(r)$

The change in the geometry of B coordination with changing composition predicted in the previous section should be evident in the partial radial distribution functions; $g_{AA}(r)$, $g_{AB}(r)$ and $g_{BB}(r)$. While $g_{AA}$ and $g_{AB}$ are dominated by the internal structure of the coordination polyhedra, the correlation function $g_{BB}$ corresponds to the correlation between polyhedra. In Figure 5 we resolve $g_{BB}$ for a $x_B = 0.25$ mixture at T = 0.4 into the contributions from $B^nA$ bonds with $1 \leq n \leq 5$. We find that individual peaks in $g_{BB}(r)$ can be quite cleanly associated with particular types of $B_nA$ bonds. The small first peak in the $A_3B$ mixture can thus be directly attributed with the relatively small number of polyhedra sharing square faces. We can also understand the anomalous temperature dependence of $g_{BB}(r)$ in which the height of the first peak *decreases* on cooling. If the triangulated coordination polyhedra such as TD and TTP are more stable at this composition, then we would expect the number of $B^4A$ bonds to decrease with the temperature.

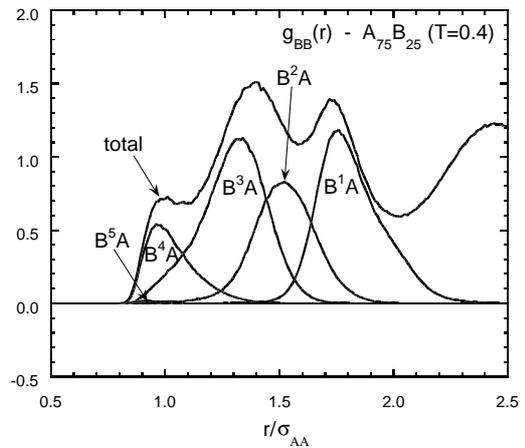

**FIGURE 5.** The contributions to the B-B radial distribution function $g_{BB}(r)$ from the $B^nA$ bonds as described in the text with $n = 1-5$ for the $x_B = 0.25$ mixture at T = 0.4. Note that the first peak is due almost exclusively to the $B^4A$ bonds and that the second peak is largely due to the $B^3A$ bonds which, at this composition, significantly exceed the $B^4A$ bonds in number.

These arguments lead us to expect a significant change in $g_{BB}$ as we increase $x_B$. We have plotted $g_{BB}$ for a range of compositions at T = 0.6 in Figure 6. Note that the relative heights of the first and second peaks undergo an inversion as we go from $x_B = 0.25$ towards the equimolar mixture. One possible explanation of this inversion is that a change in the intermediate structure as occurred, from one based on polyhedra sharing triangular faces to a structure characterised by shared square faces.

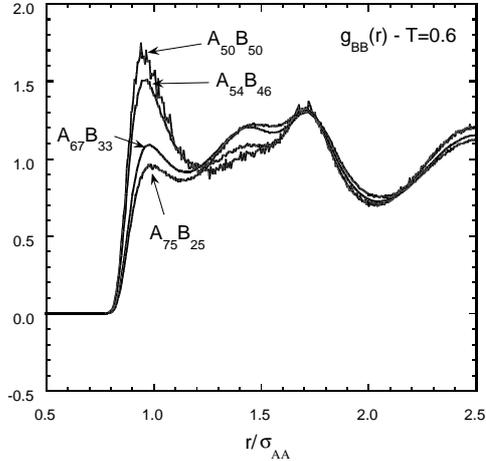

**FIGURE 6.** The B-B radial distribution function $g_{BB}(r)$ at $T = 0.6$ for the following compositions: $x_B = 0.25, 0.33, 0.46$ and $0.5$. Note the significant increase in the height of the first peak as $x_B$ is increased from 0.33 to 0.46 and the accompanying decrease in the height of the second peak.

## Crystallisation

In the previous section we noted that at compositions $x_B < 0.4$ the liquid structure, as seen through $g_{BB}$, exhibited correlations incommensurate with those in the CsCl crystal structure. Such a difference in the structures of the crystal and liquid phases would be expected to increase the interfacial free energy between the two phases. Through its inclusion in the free energy of the critical crystal nucleus in the classical theory of homogeneous nucleation, this increase in surface free energy would translate as a significant slow down in the time required for crystal growth to occur.

To explore this point, we have made a rough estimate of the minimum crystallisation time as a function of composition. The crystallisation time at a given composition and temperature is defined here as the time required after the quench for the potential energy to drop to the value midway between that of the initial disordered state and the final crystalline state. Locating the minimum such time at a given composition involves quenching the mixture to a number of different temperatures. The time required for these runs has meant that we can only perform single run at any given temperature and composition with the result that there is considerable statistical uncertainty in the estimate of the crystallisation time.

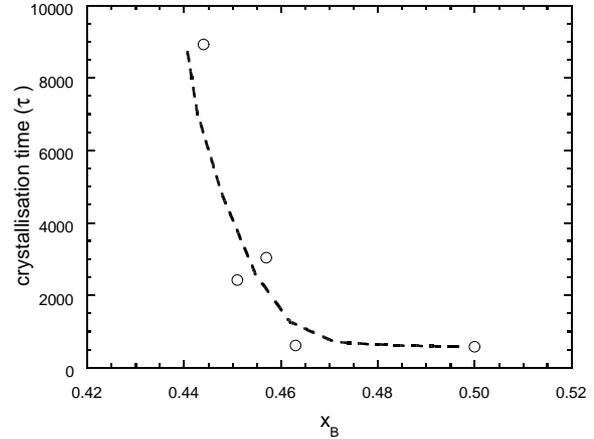

**FIGURE 7.** The minimum crystallisation time (see text for the definition) as a function of the composition $x_B$. The dashed line is included as a guide to the eye.

These minimum crystallisation times (our estimate of the 'nose' of the time-temperature transformation curves) are plotted against composition in Figure 7. Note the rapid increase in this time as the $x_B$ decreases from 0.5. For $x_B < 0.42$, we can no longer observe any sign of crystallisation over runs of $10^4 \tau$. We note that this composition lies within the range of composition identified in the previous section over which we see the inversion in the relative heights of the first and second peaks of $g_{BB}$.

## CONCLUSIONS

In this paper we have presented a preliminary account of the structure of the amorphous binary mixture introduced by Kob and Andersen. Our main results can be summarised as follows. At low temperatures we find almost all the B particles in either 8- or 9-fold coordination polyhedra made up of A particles. These polyhedra are largely triangular-faced with the result that the first peak of $g_{BB}(r)$, which reflects the number of shared square faces, is considerably smaller than the second peak. The proposal that it is the stability of these triangular-faced polyhedra that suppresses crystallisation of the CsCl phase gains support from the coincidence of the increase in magnitude of the first peak of $g_{BB}(r)$ and the onset of crystallisation as the composition $x_B$ is increased.

What remains to be explained is the extended structure and stability of the triangular-faced polyhedra for $0.2 < x_B < 0.4$ and the nature of the structural transition in the amorphous state as the composition is varied. In terms of face-sharing between adjacent polyhedra, there would seem to be some incompatibility between polyhedra with triangular faces and those with square faces which could result in an 'all-or-nothing' collective selection of one or other type of coordination. This argument neglects, however, the role of edge sharing and the stability of crystals like $Al_2Cu$ in which the local coordination exhibits both square and triangular faces. Clearly we have more work to do to understand the collective character of packing of 'soft' polyhedra.

With respect to the stability of the structure we note that recent studies of 2D glass-forming mixtures [19] have demonstrated that domains as small as a particle and its nearest neighbour shell can play the role of Tammann's 'rigid clusters' over the time scale accessible to simulations. This would suggest that fragments rather than extended structures are probably sufficient to account for the stability of simulated glasses. This does not mean that we do not need to look for more extended forms of structure but simply that we may exhaust the information that current molecular dynamics simulation methods can access.

## ACKNOWLEDGMENTS

We would like to acknowledge the support of the Australian Research Council and the Comisión Nacional de Energía Atómica of Argentina.